\documentclass[a4paper]{article}
\usepackage{qcircuit}
\usepackage{graphicx}
\usepackage[utf8x]{inputenc}
\usepackage[T1]{fontenc}
\usepackage{authblk}
\usepackage{subfig}
\usepackage[a4paper,top=3cm,bottom=2cm,left=3cm,right=3cm,marginparwidth=1.75cm]{geometry}
\input{amssym}
\usepackage[colorinlistoftodos]{todonotes}
\usepackage[colorlinks=true, allcolors=blue]{hyperref}
\usepackage{authblk}

\date{}

\title{\bf Quantum simulation of FMO complex using one-parameter semigroup of generators }
\author[1]{ M.Mahdian\thanks{mahdian@tabrizu.ac.ir}}
\author[1]{H.Davoodi Yeganeh\thanks{h.yeganeh@tabrizu.ac.ir }}

\affil[1]{Faculty of Physics, Theoretical and astrophysics department , University of Tabriz, 51665-163 Tabriz, Iran}

\begin{document}
\maketitle

\begin{abstract}
The application of open quantum systems in biological processes such as photosynthetic complexes has recently received renewed attention. In this paper, we introduce a quantum algorithm for simulation of Markovian dynamics of the Fenna-Matthew-Olson (FMO) complex that exists in photosynthesis using a ``universal set''  of a one-parameter semigroup of generators. We investigate the details of each generator that has been obtained from spectral decomposition of the Gorini-Kossakowski-Sudarshan (GKS) matrix by using linear combination and unitary conjugation. Also, we present a simple quantum circuit for the implementation of these generators.
\end{abstract}
{\bf Keyword}: Quantum simulation, FMO complex, Quantum algorithm, One-parameter semigroup.

\section{Introduction}
In recent years, the study of biological systems in which quantum dynamics are visible, and the theory of open quantum systems is applied to describe these dynamics have attracted much attention \cite{r1}. One of these biological systems in plants is a group of prokaryotes like green sulfur bacteria that utilize photosynthesis as a process to produce energetic chemical compounds by free solar energy. Harvesting of light energy and its
conversion to cellular energy currently are mainly done in photosystem complexes present in all photosynthetic organisms \cite{r2}. The photosystem is mainly constructed of two linked sections: an antenna unit includes several proteins referred to as light-harvesting complexes (LHCs) which absorb light and conduct it to the reaction center (RC). Both the LHCs and RC consist of many pigment molecules that increase the available spectrum for the photosynthesis process. After absorbing a photon, the FMO antennae complex transfers it to the RC and acts as a quantum wire between the antenna and RC \cite{r3}. The FMO complex structure is relatively simple, consisting of three monomers. Each monomer environment includes seven bacterial proteins or molecules called bacteriochlorophyll (BChl). The essential process in photosynthesis is the interaction of light with the electronic degrees of freedom of the pigment molecules, which must be studied using quantum mechanics. Furthermore, long-lived quantum coherence among the electronically excited states of the multiple pigments in the FMO complex has been shown by  2D electronic spectroscopy  \cite{pp1,pp2,pp3,pp4,pp5}. After the pigment molecule absorbs the light energy, it goes from a ground state to an excited state and also behaves like a two-level system.  Several researchers have studied the electronic excitation transfer by diverse methods such as Forster theory in weak molecular interaction limit or by Redfield master equations derived from Markov approximation in weak coupling regime between molecules and environment \cite{r3,r4,r5,r6}. Effective dynamics in the FMO complex is modeled by a Hamiltonian which describes the coherent exchange of excitations between sites and local Lindblad terms that take into account the dissipation and dephasing caused by the surrounding environment \cite{r7}.\\
It is believed that efficiently simulating quantum systems with complex many-body interactions are hard for classical computers due to the exponential growth of variables for characterizing these systems. Quantum simulation was proposed to solve such an exponential explosion problem using a
controllable quantum system as initially conjectured by Feynman \cite{r8,aa1,aa2}. The dynamical evolution of closed systems is described by unitary transformation and can be simulated directly with the quantum simulator. In the real world, all quantum systems are invariably in contact with an environment and are an open quantum system. Therefore, the dynamic evolution of these systems in the presence of decoherence and dissipation are non-unitary operations. Generally, the dynamics of an open quantum system is very complex and often used to describe the dynamics of proximity like the Born and Markov approximations are used \cite{r9}. A lot of analytical and numerical methods have been employed to simulate the dynamics of open quantum systems like composition framework for the combination
and transformation of semigroup generators, simulation of Markovian quantum dynamics by logic network and in particular simulation of arbitrary quantum channels \cite{r10,r11,r12,r13,r14,r15}. However, in these methods, there exists no universal set of non-unitary processes through which
all such processes can be simulated via sequential simulations from the universal set, but they are applicable in many problems. Rayn et al. introduce an efficient method to simulate a Markovian open quantum system, described by a one-parameter semigroup of quantum channels, which can be through sequential simulations of processes from the universal set. They used linear combination and unitary conjugation to simulate Markovian open quantum systems \cite{r16}.\\
The simulation dynamics of light-harvesting complexes are highly regarded, and a large number of various experimental and analytically studies have been conducted on them. Finding spectral density by molecular dynamics and numerical method has been studied in \cite{r17,r18,rr18}. The system dynamics simulation has been done with different platforms for implementing quantum simulators, such as two-dimensional electronic spectroscopy \cite{r20},  superconducting qubits \cite{r21}, and nuclear magnetic resonance \cite{qs1,qs2}.\\
In this paper, we use a linear combination and unitary conjugation to simulate Markovian non-unitary processes in photosynthetic FMO complex. Also, we consider constructing efficient quantum circuits based on quantum gate model for the quantum dynamics simulation subject to dissipation and dephasing environment.\\
 The remainder of the paper is organized as follows. In, Section~\ref{sec2}, we give a brief description of the affective dynamics of FMO complex and describe the universal simulation of Markovian open quantum systems and the simulation of non-unitary processes in photosynthetic FMO complex will be studying. We finally express our results in Section~\ref{sec3}.
\section{Simulation of Non-unitary Dynamics}\label{sec2}
We assume that the quantum system coupled to an environment with the Hilbert space $\mathcal{H}_S \cong \Bbb{C}^d$ (d-dimensional complex vector). The state of this system can be described by density matrix $\rho \in \mathcal{M}_d(\Bbb{C})\cong \mathcal{B}(\mathcal{H}_S),$  where $Tr[\rho]=1$ and $\mathcal{B}(\mathcal{H}_S)$ is the algebra of bounded operators on the Hilbert space. The density matrix evolves according to a Markovian quantum master equation
\begin{equation}
{\frac{d}{dt}}\rho(t)= \mathcal{L}\rho(t),
\end{equation}
where $\mathcal{L}$ is the generator of one parameter semigroup of quantum channels $\{T(t)\}$ \cite{r9}. At time $t \textgreater t_{0}$, the state of the quantum system obtained from $ \rho(t)$ = $T(t - t_0) \rho(t_0)$. In this case, we almost can write $ \mathcal{L}\rho(t)$ as follows:
\begin{equation}\label{E2}
\mathcal{L}(\rho) = i[\rho , H]+\sum_{l,k}^{d^2-1} \mathcal{A}_{l,k}(F_l\rho F^\dag_k -{1\over 2} \{ F^\dag_kF_l,\rho \}),
\end{equation}
which is known as the Gorini, Kossakowski, Sudarshan, and Lindblad (GKSL) form of the quantum Markov master equation. Note that H is, in general, a Hermitian operator, $\mathcal{A} \in \mathcal{M}_{d^2-1}(\Bbb{C})$ is the GKS matrix with the matrix elements $ \mathcal{A}_{l,k}$ and $\{F_i\}$ is basis for the space of traceless matrices in $\mathcal{M}_d(\Bbb{C})$. By diagonalization of the GKS matrix, we obtain Lindblad master equation as
\begin{equation}
\mathcal{L}(\rho) = i[\rho , H]+\sum_{k=1}^{n}\gamma_k(L_k\rho L^\dag_k -{1\over 2} \{ L^\dag_kL_k,\rho \}),
\end{equation}
where $n$ is the number of non-zero eigenvalues of $\mathcal{A}$. We begin by transforming the Lindblad master equation into the GKS form. And after, we decompose $\mathcal{A}$ into a linear combination of rank one generators through the spectral decomposition. Then, each constituent generator $ \vec{a}_k\vec{a}^\dag_k$ decomposed into the unitary conjugation of a semigroup from the universal set. See reference \cite{r11,r16} for details and proof of Theorem.
\subsection{FMO Complex}
Light-harvesting complexes seem particularly suitable as biological systems to understand quantum-mechanical effects.
Their lengthscales and energyscales are on the order where we would expect quantum-mechanical laws to apply but what remains less clear is if they can still see quantum effects such as entanglement even at physiological
temperature. So light-harvesting complexes  like the photosynthetic FMO complex exhibit such as quantum system.
The many studies such that\cite{f1,f2,f3} and so on suggest that it might be possible to observe spatial quantum correlations in
the FMO light-harvesting protein complex. Based on these experimental observations, quantum coherence across multiple chromophoric sites has been suggested
as the probable cause of the highly efficient energy transfer in
photosynthetic systems. Experimental studies of the exciton dynamics in such systems reveal rich transport dynamics consisting of short-time coherent
quantum dynamics which evolve, in the presence of noise into
an incoherent population transport which irreversibly transfers excitations to the reaction center.
The FMO complex is generally constituted of multiple chromophores which transform photons into exactions and transport to an RC. As already mentioned that efficient dynamics FMO complex express by combining Hamiltonian which describes the coherent exchange of excitations between sites, and local Lindblad terms that take into account the dephasing and dissipation caused by the external environment as non-unitary evolution  \cite{r7}.
However,in some of studies on the photosynthetic system, these quantum effects have been neglected and classical method such as the Hierarchical Equations
of Motion(HEOM) have been used to investigate this system\cite{f4,f5,f6}.\\
The exciton dynamics for the light-harvesting system (e.g., in the FMO complex) is modelled by a Markovian master equation of
the form
\begin{equation}\label{density1 }
\dot{\rho}(t)=-i[H_{sys},\rho(t)]+\mathcal{L}_{deph}(\rho)+\mathcal{L}_{diss}(\rho),
\end{equation}
which contains the coherent exchange of excitation and local Lindblad terms \cite{r7,N11}.

Since the FMO complex is composed of seven
chromophores, it should be modelled by a network of seven sites. The quantum coherent evolution of the FMO complex is determined by a
Hamiltonian of the form
\begin{equation}
H_{sys}=\sum_{i=1}^7\hbar \omega_i \sigma_i^+ \sigma_i^- +\sum_{i\neq j}^7 \hbar \nu_{ij}( \sigma_i^+ \sigma_j^-+ \sigma_i^- \sigma_j^+)
\end{equation}
where $\sigma_i^\pm$ are the raising and lowering operators that act on site $i$, $\hbar \omega_i$ are the one-site energies, and $\nu_{ij}$represents the coupling
between sites i and j.
For expressing the dynamics of the non-unitary part, assumed that the system is susceptible simultaneously to two distinct types of noise processes, a dissipative process and dephasing process. Dissipative processes pass on excitation energy with the rate $\Gamma_j$ to the environment, and the dephasing process destroys the phase coherence with the rate $\gamma_j$ of site $j^{th}$. We approached the Markovian master equation for FMO complex,  dissipative and  dephasing processes are captured with local terms, respectively, by the Lindblad super-operators as:
\begin{equation}
\mathcal{L}_{diss}(\rho) = \sum_{j=1}^7\Gamma_j(-\sigma^+_j\sigma^-_j\rho - \rho\sigma^+_j\sigma^-_j + 2\sigma^-_j\rho\sigma^+_j),
\end{equation}
\begin{equation}
\mathcal{L}_{deph}(\rho) = \sum_{j=1}^7\gamma_j(-\sigma^+_j\sigma^-_j\rho - \rho\sigma^+_j\sigma^-_j + 2\sigma^+_j\sigma^-_j\rho\sigma^+_j\sigma^-_j).
\end{equation}
Where $ \sigma^+_j =|j\rangle \langle0|$ and $ \sigma^-_j =|0\rangle \langle j|$ are raising and lowering operators for site $j $ respectively and single excitation basis $|j\rangle = |g_1\rangle \otimes ...\otimes|e_j\rangle \otimes ... \otimes |g_7\rangle $ denote one excitation
in the site $j$. Finally, the total transfer of excitation is measured by the population in the sink.

\subsection{ Dissipative Process}
For the dissipative process, we have
\begin{equation}\label{E1}
\mathcal{L}_{diss}(\rho) = \sum_{j=1}^7\Gamma_j(-\sigma^+_j\sigma^-_j\rho - \rho\sigma^+_j\sigma^-_j + 2\sigma^-_j\rho\sigma^+j).
\end{equation}
We use the fact of $\{F_k\}$ is basis for the space of traceless matrices in $\mathcal{M}_8(\Bbb{C})$ and  $\{iF_k\}$ is a basis for $su(8)$, has the following form:
\begin{equation}
\{F_i\}^7_{i=1} \equiv d^l , \qquad d^l = {1\over \sqrt{l(l+1)}}[\sum_{j=1}^l |j\rangle \langle j| -l|l+1\rangle \langle l+1|],
\end{equation}
\begin{equation}
\{F_i\}^{35}_{i=8}\equiv\{\sigma_x^{j,k}\}_{j=1}^7 |_{j<k\leq 8} , \qquad \sigma_x^{j,k} = {1\over \sqrt{2}}(|j\rangle \langle k| + |k\rangle \langle j|),
\end{equation}
\begin{equation}
\{F_i\}^{63}_{i=36}\equiv\{\sigma_y^{j,k}\}_{j=1}^7 |_{j<k\leq 8} , \qquad \sigma_y^{j,k} = {1\over \sqrt{2}}(-i|j\rangle \langle k| + i|k\rangle \langle j|).
\end{equation}
Now, we can write $ \sigma^+_j $ and $ \sigma^-_j $ as below
\begin{equation}
\sigma^+_j={1\over \sqrt{2}} (F_{j+7} -i F_{j+35}) \ , \quad
\sigma^-_j={1\over \sqrt{2}} (F_{j+7} +i F_{j+35}).
\end{equation}
So, by substituting in Eq.(\ref{E1}) the GKS matrix can be obtained and nonvanishing elements are
$$a_{8,8}=\Gamma_1 \quad a_{36,36}=\Gamma_1 \quad a_{8,36}=-i\Gamma_1 \quad a_{36,8}=i\Gamma_1 $$
$$a_{9,9}=\Gamma_2 \quad a_{37,37}=\Gamma_2 \quad a_{9,37}=-i\Gamma_2 \quad a_{37,9}=i\Gamma_2 $$
$$a_{10,10}=\Gamma_3 \quad a_{38,38}=\Gamma_3 \quad a_{10,38}=-i\Gamma_3 \quad a_{38,10}=i\Gamma_3 $$
$$a_{11,11}=\Gamma_4 \quad a_{39,39}=\Gamma_4 \quad a_{11,39}=-i\Gamma_4 \quad a_{39,11}=i\Gamma_4 $$
$$a_{12,12}=\Gamma_5 \quad a_{40,40}=\Gamma_5 \quad a_{12,40}=-i\Gamma_5 \quad a_{40,12}=i\Gamma_5 $$
$$a_{13,13}=\Gamma_6 \quad a_{41,41}=\Gamma_6 \quad a_{13,46}=-i\Gamma_6 \quad a_{46,13}=i\Gamma_5 $$
$$a_{14,14}=\Gamma_7 \quad a_{42,42}=\Gamma_7 \quad a_{14,42}=-i\Gamma_7 \quad a_{42,14}=i\Gamma_7. $$
Next, we decompose $\mathcal{A}$ as:
\begin{equation}
\mathcal{A}=\sum_{k=1}^{7} \lambda_k \vec{a}_k\vec{a}^\dag_k,
\end{equation}
where $\lambda_k=2 \Gamma_k$ and $a_k$ have Nonvanishing elements $a_{k+7}={- i{1 \over \sqrt{2}}} $ and $ a_{k+35}={1 \over \sqrt{2}}$.
Now, each generator $\vec{a}_k\vec{a}^\dag_k$ of the linear combination should be decomposed into the unitary conjugation of a semigroup from the universal set. In general form, we can write
\begin{equation}
e^{i{\psi _k}}\vec{a}_k = \cos(\theta_k)\hat{a}^R_k +i\sin(\theta_k)\hat{a}^I_k,
\end{equation}
where $\hat{a}^R_k$ and $\hat{a}^I_k$ are real and imaginary part of $a_k$, respectively and $\psi$ is the phase transformation. If $\hat{a}^R_k.\hat{a}^I_k =0$ and $|\hat{a}^R_k|=|\hat{a}^I_k|=1$ then the phase transformation is $\psi=0$ and $\theta_k \in [0,\pi/4]$.\\
In our problem, we find, after simplifying: $\psi_k=0$ , $\theta_k = \pi /4$
, $ \vec{\alpha}^R_k = 0$ and $ \vec{\alpha}_k^I = (a_1,................,a_{35})^T$ by $ a_i= \pi/2 , \ i=1,2,....34$ and $ a_{35}=3\pi/2$
for $k=1,2,....7$.
So, we can implement
\begin{equation}
\vec{a}_k\vec{a}^\dag_k =G_{U^{(k)}} [A^{(k)}(\theta_k,\vec{\alpha}_k^R,\vec{\alpha}_k^I )]G_{U^{(k)}}^T,
\end{equation}
where $A^{(k)}(\theta_k,\vec{\alpha}_k^R,\vec{\alpha}_k^I )$ represents an element of universal set of semigroup generators and $G_{U^{(k)}} \in SO(3)$ is adjoint representation of SU(2) (see \cite{r11} for details). By considering $ U^{(k)}=U_k^{{(1)}\dag} = U_n\otimes U_{lm}$ that $ U_n$ is single unitary operation $ R_y(-{\pi \over 2})$ and $U_{lm}$ is two-qubit gate. Furthermore, if $\mathcal{L}_k$ is generator of a Markovian semigroup,
we can simulate any channel $T_k(t) = exp(t\mathcal{L}_{\vec{a}_k\vec{a}^\dag_k})$ from the semigroup generated by $ \vec{a}_k\vec{a}^\dag_k$,
\begin{equation}
T_k(t)(\rho)=U^{(k) \dag}(T_{A^{(k)}}(t)[U^{(k)}\rho U^{(k)\dag}])U^{(k)},
\end{equation}
where $T_{A^{(k)}}(t)=\exp(t\mathcal{L}_{A^(k)})$. We drive this equation for semigroup generated
by $\vec{a}_1\vec{a}^\dag_1$ directly. For $\vec{a_1}$, we obtain $\hat{a}^R_1 =|36\rangle $ , $\hat{a}^I_1=-|8\rangle$. \\
The next step is finding $\tilde{a}_1^R$ and $\tilde{a}_1^I$, for this work we use of map $f: su(8) \rightarrow \Bbb{R}^{63}$ that define as $f(iF) =|j\rangle$. If define $ \hat{A}_1^R \equiv f^{-1}(\hat{a}^R_1)$, we have
\begin{equation}
\hat{A}_1^R= iF_{36},
\end{equation}
by using of the matrix $U_1^{(1)}$, the matrix $\hat{A}_1^R$ can be diagonalized as
\begin{equation}
U_1^{(1)}=
\pmatrix{
{1\over \sqrt{2}} &{1\over \sqrt{2}}&0&0&0&0&0&0 \cr {-1\over \sqrt{2}} &{1\over \sqrt{2}}&0&0&0&0&0&0 \cr 0 &0&0&0&0&0&0&1 \cr 0 &0&0&0&0&0&1&0 \cr 0 &0&0&0&0&1&0&0 \cr 0 &0&0&0&1&0&0&0 \cr 0 &0&0&1&0&0&0&0 \cr 0 &0&1&0&0&0&0&0
},
\end{equation}

\begin{equation}
\tilde{A}^R_{d,1} \equiv U_1^{(1)} \hat{A}_1^R U_1^{{(1)}\dag} = iF_1 .
\end{equation}
For imaginary part
\begin{equation}
\hat{A}_1^I= - iF_{8},
\end{equation}
and
\begin{equation}
{A}^I_{1} \equiv U_1^{(1)} \hat{A}_1^I U_1^{{(1)}\dag} = -iF_{36} \equiv \tilde{A}^I_{1},
\end{equation}
we need not to find $U_2^{(2)}$ because of ${A}^I_{1}$ have desired form. So
\begin{equation}
f(\tilde{A}^R_{d,1}) = |1\rangle , \quad f(\tilde{A}^I_{1})=- |36\rangle .
\end{equation}
If we define $ \tilde{a}^R_1 \equiv f(\tilde{A}^R_{d,1}) $ and $ \tilde{a}^I_1\equiv f(\tilde{A}^I_{1}) $ we haven't second unitary transformation because $\tilde{a}^R_1$ and $\tilde{a}^I_1$ have desired form. So, we can implement
\begin{equation}
\vec{a}_1\vec{a}^\dag_1 =G_{U^{(1)}} [A^{(1)}(\theta_1,\vec{\alpha}_1^R,\vec{\alpha}_1^I )]G_{U^{(1)}}^T,
\end{equation}
where $A^{(1)}(\theta_1,\vec{\alpha}_1^R,\vec{\alpha}_1^I )$ is an element of the universal set of semigroup generators, by $\theta_1 =\pi/4$ ,$\vec{\alpha}_1^R=0$ , $\vec{\alpha}^R_1 = 0$ and $ \vec{\alpha}_1^I = (a_1,................,a_{35})^T$ by $ a_i= \pi/2 , \ i=1,2,....34$ and $ a_{35}=3\pi/2$. Furthermore if $\mathcal{L}_1$ is generator of a Markovian semigroup we can simulate any channel $T_1(t) = exp(t\mathcal{L}_{\vec{a}_1\vec{a}^\dag_1})$ from the semigroup
generated by $ \vec{a}_1\vec{a}^\dag_1$,
\begin{equation}
T_1(t)(\rho)=U^{(1) \dag}(T_{A^{(1)}}(t)[U^{(1)}\rho U^{(1)\dag}])U^{(1)},
\end{equation}
where $T_{A^{(1)}}(t)=\exp(t\mathcal{L}_{A^(1)})$. Now, we are designing the quantum circuit for implement of $U^{(i)}_1$. At first we obtain quantum circuit for implement $U^{(1)}_1$ and for other $U^{(i)}_1$ similarly circuit can be designed. By finding action of unitary operation $U^{(1)}_1$ on the $|q_1,q_2,q_3\rangle$ where three qubits space bases, we obtain the following conditions:
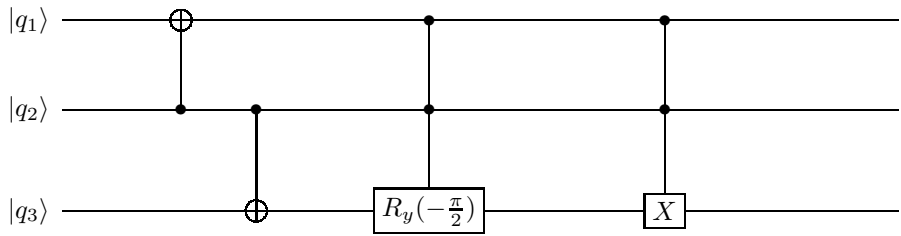
\begin{figure}[b!]
\centerline{
\Qcircuit @C=2em @R=1cm { \lstick{|q_1\rangle} & \qw & \targ \qw & \qw & \qw &\ctrl{2} & \qw & \qw &\ctrl{2} & \qw & \qw & \qw & \qw \\ \lstick{|q_2\rangle} & \qw & \ctrl{-1} & \ctrl{1} & \qw & \ctrl{-1}& \qw & \qw & \ctrl{1} & \qw & \qw & \qw & \qw \\ \lstick{|q_3\rangle} & \qw &\qw & \targ & \qw & \gate{ R_y(-{\pi \over 2})} & \qw & \qw & \gate{X} & \qw & \qw & \qw & \qw
}
}
\caption{Quantum circuit for implementing of $U_1^{(1)- diss}$}
\end{figure}
\begin{enumerate}
\item If the second qubit was in state |$1\rangle$, apply controlled-NOT (CNOT) to gate $(X)$ on the first and third qubit.
\item If first and second qubit was in state $|0\rangle$, apply the rotation gate $(Y)$ on the third qubit.
\item If the first qubit were in state $|1\rangle$ and second qubit in state $|0\rangle$, apply CNOT to gate $(X)$ on the third qubit.
\end{enumerate}
Given the above conditions, we need two CNOT gate, a single qubit gate($R_y(\frac{-\pi}{2}))$ and an X gate.
Quantum circuit for implement of $U^{(1)}_1$ shown in Fig.1\\

\subsection{Dephasing Process}
Similarly, in the previous section, we obtain the GKS matrix for this process and decompose it. So, we have
\begin{equation}
\mathcal{A}=\sum_{k=1}^{7} \lambda_k \vec{a}_k\vec{a}^\dag_k,
\end{equation}
with $ \lambda_k = 4\gamma _k$ for $k=2,3,..,7$ and $\lambda_1=4\sqrt{2}\gamma _1$.
$\vec{a_1}$ have Nonvanishing elements $a_{10}= {-i(1+\sqrt{2})\over \sqrt{4+2 \sqrt{2}}}\ , \ a_{38}= {1\over \sqrt{4+2 \sqrt{2}}}.$
Nonvanishing elements of $\vec{a_2}$ are $a_{16}= {1\over \sqrt{5}}( -2i)\ , \ a_{47}= {1\over \sqrt{5}}$ and for $\vec{a_3}$: $a_{12}= {1\over \sqrt{5}}(-2i)\ , \ a_{40}= {1\over \sqrt{5}}$. As the same way Nonvanishing elements of $\vec{a_4}$ ,$\vec{a_5}$,$\vec{a_6}$,$\vec{a_7}$ are
$( a_{20}= {1\over \sqrt{2}}i \ , a_{47}={1\over \sqrt{2}})$,$( a_{19}= {1\over \sqrt{2}}i \ , a_{46}={1\over \sqrt{2}})$,$( a_{16}= {-1\over \sqrt{2}}i \ , a_{45}={1\over \sqrt{2}})$,$( a_{11}= {-1\over \sqrt{2}}i \ , a_{39}={1\over \sqrt{2}})$, respectively.
$\psi_k = \pi/2$ for $k=1,2,3$ and equal with zero for $k=4,...,7$. $\theta_1= \arccos({1+\sqrt{2} \over \sqrt{4+2\sqrt{2}}})$ , $ \theta_2=\theta_3=\arccos({2\over \sqrt{5}})$ and $\theta_k=\pi/4$ for $ k=4,...,7$.\\
Furthermore, we can obtain $\vec{\alpha}^R_{1,7} = 0$ and $ \vec{\alpha}_{1,7}^I = (a_1,................,a_{35})^T$ by $ a_i= \pi/2 , \ i=1,2,....34$ and $ a_{35}=3\pi/2$ for $\vec{a}_{2,3}$ can be written $\vec{\alpha}_{2,3}^R=0 ,\ \vec{\alpha}_{2,3}^I=\pi$, and for $\vec{ a}_{4,5,6}$ we obtain  $ \vec{\alpha}_{4,5,6}^R= \vec{\alpha}_{4,5,6}^I=\pi.$\\
Now, we consider semigroup generated by $\vec{a}_1\vec{a}^\dag_1$ and decompose it into the unitary conjugation of a semigroup from the universal set.
We begin by $\hat{a}^R_1 =|10\rangle$ , $\hat{a}^I_1= |38\rangle$ and
$$ f^{-1}(\hat{a}^R_1)=f^{-1}(|10\rangle) = iF_{10} = \hat{A}^R_1,$$
now by using of $U^{(1)}_1$ we can diagnose the matrix $\hat{A}^R_1$ .
\begin{equation}
U_1^{(1)}=
\pmatrix{
{1\over \sqrt{2}} &0&0&{1\over \sqrt{2}}&0&0&0&0 \cr {-1\over \sqrt{2}} &0&0&{1\over \sqrt{2}}&0&0&0&0 \cr 0 &0&0&0&0&0&0&1 \cr 0 &0&0&0&0&0&1&0 \cr 0 &0&0&0&0&1&0&0 \cr 0 &0&0&0&1&0&0&0 \cr 0 &0&0&1&0&0&0&0 \cr 0 &0&1&0&0&0&0&0
},
\end{equation}
Then, we'll have
$$U_1^{(1)}\hat{A}^R_1U_1^{(1) \dag}=iF_1=\tilde{A}^R_{d,1},$$
for an imaginary part
$$ f^{-1}(\hat{a}^I_1)=f^{-1}(|38\rangle) = iF_{38} = \hat{A}^R_1 ,$$
and
$$U_1^{(1)}\hat{A}^I_1U_1^{(1) \dag}=iF_{36}=\tilde{A}^I_{1},$$
because $\hat{A}^I_1$ is desired form, no need to find a matrix $U_2^{(1)}$.
If we define $\tilde{a}^R_1 \equiv f(\tilde{A}^R_{d,1})=|1\rangle$ and $\tilde{a}^I_1 \equiv f(\tilde{a}^I_{1})=|36\rangle$.
We need not to second unitary transformation because $\tilde{a}^R_1$ and $\tilde{a}^I_{1}$ are have the desired form.
By consider $ U^{(1)}=U_1^{{(1)}\dag}$ and similar to the previous one
\begin{equation}
\vec{a}_1\vec{a}^\dag_1 =G_{U^{(1)}} [A^{(1)}(\theta_1,\vec{\alpha}_1^R,\vec{\alpha}_1^I )]G_{U^{(1)}}^T,
\end{equation}
where $A^{(1)}(\theta_1,\vec{\alpha}_1^R,\vec{\alpha}_1^I )$ an element of the universal set
of semigroup generators, by $\theta_1 =\arccos({1+\sqrt{2} \over \sqrt{4+2\sqrt{2}}})$ ,$\vec{\alpha}_1^R=0$ and $ \vec{\alpha}_1^I = (a_1,................,a_{35})^T$ by $ a_i= \pi/2 , \ i=1,2,....34$ and $ a_{35}=3\pi/2$.
Furthermore if $\mathcal{L}_1$ is generator of a Markovian semigroup we can simulate any channel $T_1(t) = exp(t\mathcal{L}_{\vec{a}_1\vec{a}^\dag_1})$ from the semigroup
generated by $ \vec{a}_1\vec{a}^\dag_1$,
\begin{equation}
T_1(t)(\rho)=U^{(1) \dag}(T_{A^{(1)}}(t)[U^{(1)}\rho U^{(1)\dag}])U^{(1)},
\end{equation}
where $T_{A^{(1)}}(t)=\exp(t\mathcal{L}_{A^{(1)}})$. Note in any case must be calculated $U_i^{(1)}$ and $ U_i^{(2)}$ for $i=2,...,6$ except in case $\vec{a}_{1,7}$ that their calculation is straightforward. Now we design quantum circuit to implement of $U_1^{(1)}$. We drive the following conditions by finding action of $U_1^{(1)}$ on three qubits space bases.

\begin{enumerate}
\item If the first and second qubit was in state $|0\rangle$ and $|1\rangle$, apply the rotation gate $(Y)$ on the third qubit.
\item If the first qubit was in state $|1\rangle$ and second qubit in state $|0\rangle$, apply CNOT to gate $(X)$ on the third qubit.
\item If the second qubit was in state |$1\rangle$, apply CNOT gate to $(X)$ on the first and third qubit.
\end{enumerate}
Furthermore, via two CNOT and single-qubit gates $ R_y(-{\pi \over 2})$ and a $X$ gate, unitary operation $U_1^{(1)}$ can be implement. Quantum circuit of $U_1^{(1)}$ shown in Fig.2. For other $U_i^{(1)}$ quantum circuit can be design analogous $U_1^{(1)}$.
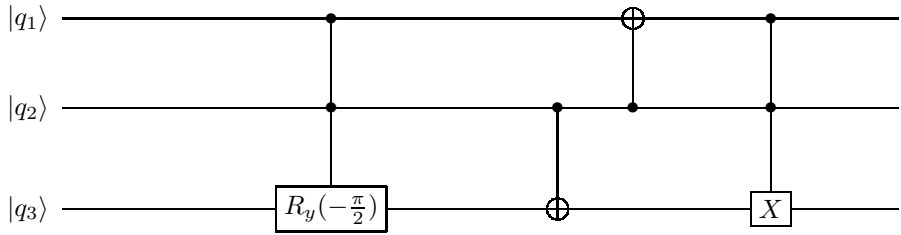
\begin{figure}[h!]
\centerline{
\Qcircuit @C=2em @R=1cm { \lstick{|q_1\rangle} & \qw & \qw & \qw & \ctrl{1} & \qw & \qw & \qw & \targ & \qw & \ctrl{2} & \qw & \qw & \\ \lstick{|q_2\rangle} & \qw & \qw & \qw & \ctrl{1}& \qw & \qw &\ctrl{1} & \ctrl{-1} & \qw & \ctrl{1} & \qw & \qw & \\ \lstick{|q_3\rangle} & \qw &\qw & \qw & \gate{ R_y(-{\pi \over 2})} & \qw & \qw & \targ & \qw & \qw &\gate{X} &\qw & \qw &
}
}
\caption{Quantum circuit for implementing of $U_1^{(1)-deph}$.}
\end{figure}

\section{Conclusion}\label{sec3}
In this paper, we have investigated the universal simulation of Markovian dynamics of the FMO complex. At first, we have transformed the Lindblad master
equation into the GKS form for non-unitary processes in the FMO complex. Next, decomposed the GKS matrix into the linear combination of rank one generators through spectral decomposition. Then each constituent generator
$ \vec{a}_k\vec{a}^\dag_k$ decomposed into the unitary conjugation of a semigroup from the universal set.
Finally, the quantum circuit had designed for implementing a unitary matrix that applied for simulation of the structure generators.

\bibliographystyle{unsrt}
\bibliography{re2.tex}

\end{document}